\documentclass[epj]{svjour}
\usepackage{epsfig,float}

\begin{document}

\title{Enhanced excitation of Giant Pairing Vibrations
in heavy-ion reactions induced by weakly-bound projectiles.}

\author{L. Fortunato\inst{1}\thanks{\email{fortunat@pd.infn.it}}
 \and W. von Oertzen \inst{2}\and  H.M. Sofia \inst{3} \and 
A. Vitturi \inst{1}}
\institute{1) Dipartimento di Fisica and INFN, Padova, Italy \and
2) Hahn Meitner Institute, Berlin, Germany \and
3) Comisi\'on Nacional de Energ\'{\i}a At\'omica and CONICET, 
Buenos Aires, Argentina}

\titlerunning{Excitation of Giant Pairing Vibration.}
\authorrunning{L.Fortunato et al.}

\abstract{
The use of radioactive ion beams is shown to offer the possibility
to study collective pairing states at high excitation energy,
which are not usually accessible with stable projectiles because of 
large energy mismatch. In the case of two-neutron stripping
reactions induced by $^{6}$He, 
we predict a population of the Giant Pairing
Vibration in $^{208}$Pb or $^{116}$Sn with cross
sections of the order of a millibarn, dominating over the
mismatched transition to the ground state.}

\maketitle

\section{Introduction.}

Large efforts have been recently dedicated to the study of
different aspects of reaction mechanism in collisions induced
by weakly-bound radioactive beams.  The long tails
of the one-particle transfer form factors due to the
weak binding, associated with the possibility of unusual
behaviour of pairing interaction in diluted systems, has raised
novel interest in the possibility of studying the pair field via
two-particle
transfer processes with unstable beams \cite{VOV}. On the other hand,
in transfer reactions induced by weakly bound projectiles 
on stable targets, the Q-values for the low-lying states will
be very large (typically of the order of 10-15 MeV
for the ($^{6}$He,$^{4}$He) stripping reaction).  This will strongly 
hinder these processes
for reactions where the semi-classical optimum matching 
conditions apply, as it is the case of bombarding energies 
around the Coulomb barrier on heavy target nuclei. Higher bombarding 
energies, where the matching conditions are less stringent, may
on the other hand not be suitable because of large break-up 
cross sections.  The same matching conditions will favour 
instead the population of highly excited states, as the 
Giant Pairing Vibrations (GPV), and the use of Radioactive
Ion Beams (RIB)
 may therefore become instrumental in offering
the opportunity of studying nuclear structure aspects that are not
usually accessible with stable projectiles.  
These Giant Pairing Vibrations are in fact predicted \cite{BB}
to have strong collective features, but their observation
may have so far failed \cite{WvO} because of large mismatch in reactions 
induced by protons or tritons, at variance to the case of
the low-lying pairing vibrations, which  
have been intensively and successfully studied around 
closed shell nuclei in two-particle transfer reactions \cite{BM}. 
All these 0$^+$
states are associated with vibrations of
the Fermi surface and are described
in a microscopic basis of the shell model as correlated 
two particle- two hole
states.  In the case of the Giant Pairing Vibrations the excitation
involves the promotion of a pair of particles (or holes) in the next
major shell (hence an excitation energy around 2$\hbar\omega$) and
is expected to display a collective pairing strength comparable
with the low-lying vibrations.  Also in the case of superfluid
systems in an open shell the system is expected to display a collective 
high-lying state, that in this case collects its strength from the
unperturbed two-quasiparticle 0$^+$ states with energy 2$\hbar\omega$.
To investigate this possibility we made estimates of cross sections
to the Giant Pairing Vibrations in two-particle transfer reactions,
comparing the cases of bound or weakly-bound projectiles.
As examples we have considered the case of ($^{14}$C,$^{12}$C), from 
one side, and the case of ($^{6}$He,$^{4}$He) as representative of 
a reaction induced by a weakly-bound ion.  As targets, we have chosen
the popular cases of the lead and tin regions (so considering
both ``normal'' and ``superfluid'' nuclei).
To perform the calculation,
we will first evaluate the response to the pairing 
operator in the RPA, including both the 
low-lying and high-lying pairing vibrations.  As a following step 
we will then construct two-neutron transfer form factors, using
the ``macroscopic'' model for pair-transfer processes. Finally,
estimates of cross sections will be given using standard DWBA
techniques. As we will see, in the case of the stripping reaction
induced by $^{6}$He, the population of the GPV is expected to display
cross sections of the order of a millibarn, dominating over the
mismatched transition to the ground state.

The paper is organized as follows.
In the next section we discuss the theoretical formalism used for normal
and for superfluid nuclei. In section 3 we recall the basics
aspects of the macroscopic form factors for two-particle transfer reactions 
and  in section 4 we display the results of calculations for the paradigmatic examples of 
$^{208}$Pb and $^{116}$Sn.

\section{The pairing response and the Giant Pairing Vibration.}

A simple way of displaying the amount of pairing correlations is in terms of 
the pair transfer transition densities \cite{Dal}. These are defined as the matrix 
element of the pair density operator connecting the ground state in nucleus 
$A$ with the generic $0^{+}$ state $|n\rangle$ in nucleus $A\pm 2$, namely
\begin{equation}
\delta\rho_{P} (r)= \langle n| \hat{\rho}_{P} |0\rangle,
\end{equation}
where the generalized density operator is given by
\begin{equation}
\hat{\rho}_{P} (r)= \sum_{\alpha}{\sqrt{2j+1}\over{4\pi}} R_{\alpha}(r)
R_{\alpha}(r)
([a_{\alpha}^{\dagger}a_{\alpha}^{\dagger}]_{00} + [a_{\alpha}a_{\alpha}]_{00}).
\end{equation}
Here $ R_{\alpha}(r)$ are the radial wave functions of the 
$\alpha = \{ nlj \}$ level and the sum runs over both particle and hole levels.
The pair transfer strength to each final state can be obtained from the 
corresponding pair transfer transition density by simple quadrature, namely
\begin{equation}
\beta_{P} = \int 4\pi r^{2} \delta\rho_{P} dr.
\end{equation} 

For normal systems around closed shell the strong L=0 transition follows a 
vibrational scheme, 
where the correlated pair of fermions (pairing phonon) change by one \cite{RPA}.
In this case, there are two types of phonons associated with the stripping and 
pick-up  reactions. The two-particle collective state is called "addition" 
pairing phonon while the two-holes correlated state is known as "removal" 
pairing phonon. From a microscopic point of view the two kind of 
phonons, corresponding to the $(A\pm 2)$ nuclei 
can be described in terms of the two-particle (two-hole) states of the
Tamm Dancoff Approximation(TDA) or in a better way by a Random Phase 
Approximation (RPA). We start from a hamiltonian with a Monopole Pairing 
interaction:
\begin{equation}
H=\sum_{j} \epsilon_{j} a^{\dagger}_{j}a_{j} - G4\pi P^{\dagger}P,
\label{p1}
\end{equation}
where
\begin{equation}
P^{\dagger}=\sum_{j_{1} \le j_{2}} {M(j_{1},j_{2}) \over 
\sqrt{1+\delta_{j_1j_2} }}
\left[ a^{\dagger}_{j_{1}}a^{\dagger}_{j_{2}}\right]_{00}.
\label{p2}
\end{equation}
Here the $a^{\dagger}_{j}$ creates a particle in an orbital $j$, where $j$ 
stands for all the needed quantum numbers of the level.
The constant $G$ is the strength of the pairing interaction and the 
coefficients 
$M(j_{1},j_{2})$ are:
\begin{equation}
M(j_{1},j_{2})={\langle j_{1}||f(r) Y_{00}(\theta,\phi)||j_{2}\rangle\over 
\sqrt{1+\delta_{j_1j_2} }},
\label{p3}
\end{equation}
where the detailed radial dependence of $f(r)$ is taken to
be of the form $r^L$ and in our case is a constant since we are dealing 
only with $L=0$ states.
The Pairing phonons are defined for closed shell nuclei as:
\begin{equation}
\begin{array}{lll}
|n,2p\rangle = \Gamma_{n,2p}^{\dagger}|0\rangle =&~&~ \\
\qquad =\sum_{k} X_{n}(k) [a^{\dagger}_k a^{\dagger}_k]_{00} + 
\sum_{i} Y_{n}(i) [a_i a_i]_{00}&~&~\\
&&\nonumber\\
|n,2h\rangle = \Gamma_{n,2h}^{\dagger}|0\rangle = &~&~ \nonumber\\
\qquad = \sum_{i} X_{n}(i) [a^{\dagger}_i a^{\dagger}_i ]_{00} +  
\sum_{k}Y_{n}(i) [a_k a_k ]_{00}&~&~ ,\\
\label{p4}
\end{array}
\end{equation}
where $k (i)$ stands for levels above (below) the Fermi level. We will use  
$j$ for any of both cases.  $X_n$ and $Y_n$ are the forward and backward 
amplitudes.
Within this model the pair transfer strength associated with each RPA state is
microscopically given by
\begin{equation}
\beta_{Pn} = \sum_{j} \sqrt{2j+1} [X_{n}(j) + Y_{n}(j)] .
\label{p5}
\end{equation}

In Fig. 1a we display the predicted pairing response in the case of $^{206}$Pb, 
namely two-neutron holes with respect to the double magic $^{208}$Pb.
The set of single-particle levels that has been used in the RPA calculation, 
was obtained using the 
spherical harmonic oscillator levels with corrections due to the 
centrifugal and spin-orbit interactions \cite{MJ}
\begin{equation}
{E\over{\hbar\omega}}= N+{3\over 2} - \mu \left(l(l+1)-{{N(N+3)}
\over 2}\right) + K 
\end{equation}
$$ K = \left\{  \begin{array}{ll} 
-\kappa l & \mbox{  for  } j=l+1/2 \\
-\kappa(l+1) & \mbox{  for  } j=l-1/2
\end{array} \right. ,$$
where $\hbar\omega=41 A^{-{1\over 3}}$, $A$ is the mass number of the
nucleus, $N$ is the principal quantum number and $ j, l$ are the
total and orbital angular momentum quantum numbers, respectively. The
quantities $\kappa$ and $\mu$ are parameters chosen to obtain the best 
fit for each nucleus \cite{Da}.
We have included in the calculation all  the single-particle levels starting 
from $N=0$ up to 10. 
This set is expected to be good enough for our calculation of the Giant 
Pairing Resonance, except for the levels around the Fermi surface. 
In the lead region we prefer to use experimental values for the 
shells just above and below the Fermi surface \cite{GLN}.
The Figure shows, in addition to the strong collectivity associated with the 
ground state transition, a strong collective state with about half of the 
g.s. strength at high excitation energy, around 16 MeV, which can be 
interpreted as the Giant Pairing Vibration. Similar situation is shown in 
Fig. 1b for the corresponding two-neutron addition states in the $^{210}$Pb.
Again one may interpret the strength at about 12 MeV as associated 
with the giant mode.
\begin{figure}
\resizebox{1.0\hsize}{!}{\includegraphics*{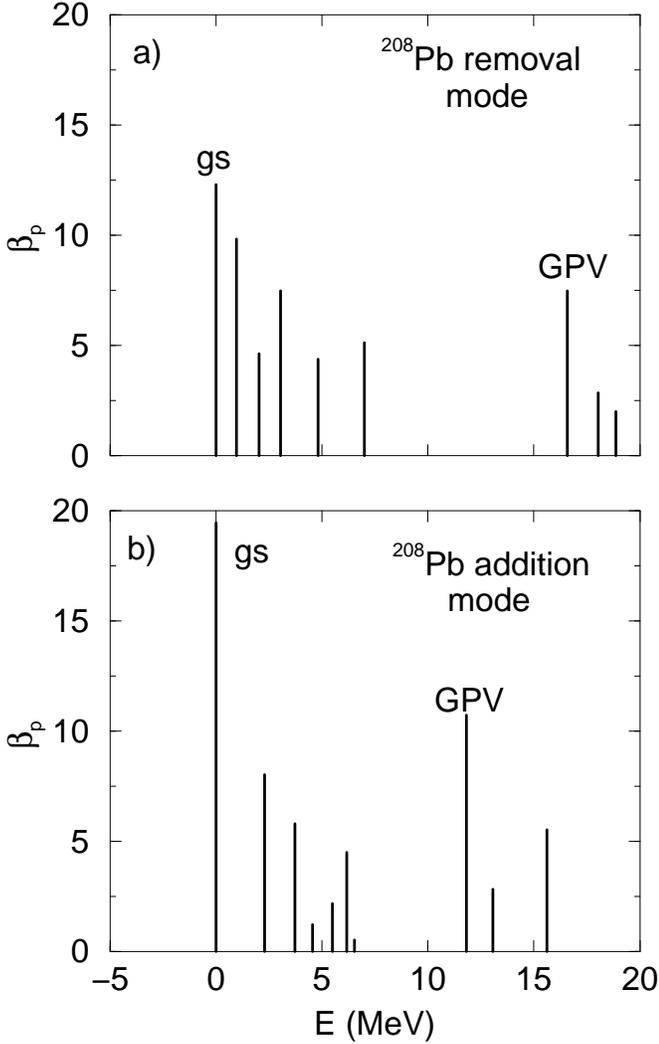}}
\caption{ Pairing response for the removal (upper panel) and addition 
(lower panel) mode in $^{208}$Pb. The ground state transition and the candidate 
for GPV are evidenced.}
\end{figure}

We consider now the case of superfluid spherical-nuclei. In this case we make a 
BCS transformation of the hamiltonian defined in Eq. [\ref{p1}] changing from 
particle to quasiparticle operators, introducing the usual occupation 
parameters. We start from a single-quasiparticle  
Hamiltonian plus a two-quasiparticle interaction corresponding to the residual
$H_{22} + H_{40}$ of the pairing force
\begin{eqnarray}
H &=& \sum_{j} E_{j} \alpha^{\dagger}_{j} \alpha_{j} \nonumber \\
&+& 2 \pi G \sum_{j_{1}j_{2}} M(j_1,j_1) M(j_2,j_2) \cdot \nonumber\\
& \cdot \Bigl\{ ~&(U^{2}_{j_{1}} U^{2}_{j_{2}} + V^{2}_{j_{1}} V^{2}_{j_{2}})
[\alpha^{\dagger}_{j_{1}}\alpha^{\dagger}_{j_{1}}]_{00} [\alpha_{j_{2}}
\alpha_{j_{2}}]_{00}\nonumber \\ 
&-&  U^{2}_{j_{1}} V^{2}_{j_{2}}
[\alpha^{\dagger}_{j_{1}}\alpha^{\dagger}_{j_{1}}]_{00}
[\alpha^{\dagger}_{j_{2}}\alpha^{\dagger}_{j_{2}}]_{00}  \nonumber \\
&-& V^{2}_{j_{1}} U^{2}_{j_{2}}[\alpha_{j_{1}}\alpha_{j_{1}}]_{00} 
[\alpha_{j_{2}}\alpha_{j_{2}}]_{00}\Bigr\} ,
\label{p7}
\end{eqnarray}
where
\begin{eqnarray}
\alpha^{\dagger}_{j} &=& U_{j} a^{\dagger}_{j} - V_{j} a_{\bar j}\\
U^{2}_{j} &=& {1\over {2}} \left(1 + {{\tilde \epsilon_{j}}\over E_{j}}\right)\\
V^{2}_{j} &=& {1\over 2} \left(1 - {{\tilde \epsilon_{j}}\over E_{j}}\right).
\label{p8}
\end{eqnarray}
The energies $E_{j} = \sqrt{{\tilde \epsilon}_{j}^{2} + \Delta^{2}}$ are the 
quasi-particle energies, and ${\tilde \epsilon_{j}} = \epsilon -\lambda$ 
are the single-particle energies with respect to the chemical potential
$\lambda$  and $\Delta$ is the BCS gap. 
As usual we have defined $a_{\bar j} \equiv a_{\bar{jm}}=(-1)^{j-m} a_{j,-m}$.
 
For superfluid systems the addition and removal RPA phonons cannot be treated 
separately. The dispersion relation, that relates the strength of the 
interaction with the energy-roots of the RPA, becomes a two by two determinant.
From the RPA equations:
\begin{equation}
\Gamma^{\dagger}_{n} ~=~ \sum_{j} \left( X_{n}(j) [\alpha^{\dagger}_{j} 
\alpha^{\dagger}_{j}]_{00} + Y_{n}(j)[\alpha_{j}\alpha_{j}]_{00}\right)
\label{p9uno}
\end{equation}
\begin{equation}
 \left[ H,\Gamma^{\dagger}_{n}\right] ~=~ \omega_{n} \Gamma^{\dagger}_{n},
\label{p9}
\end{equation}
we can obtain the following factors
\begin{eqnarray}
x &=& \sum_{j_{1} \le j_{2}} |M(j_{1}j_{2})|^{2} \left[\frac{U^{2}_{j_{1}} 
U^{2}_{j_{2}}}{E_{j_{1}}+E_{j_{2}}-\omega_{n}} + \frac{V^{2}_{j_{1}} 
V^{2}_{j_{2}}}{E_{j_{1}}+E_{j_{2}}+\omega_{n}}\right]\nonumber \\
~\\
y &=& \sum_{j_{1} \le j_{2}} |M(j_{1}j_{2})|^{2} \left[\frac{V^{2}_{j_{1}}
V^{2}_{j_{2}}}{E_{j_{1}}+E_{j_{2}}-\omega_{n}} + \frac{U^{2}_{j_{1}}
U^{2}_{j_{2}}}{E_{j_{1}}+E_{j_{2}}+\omega_{n}}\right] \nonumber \\
~\\
z &=& \sum_{j_{1} \le j_{2}} |M(j_{1}j_{2})|^{2} (U_{j_{1}}V_{j_{2}}
U_{j_{2}}V_{j_{1}}) \nonumber \\
~&~& \left[\frac{1}{E_{j_{1}}+E_{j_{2}}-\omega_{n}} + 
\frac{1}{E_{j_{1}}+E_{j_{2}}+\omega_{n}}\right] ,
\label{p10}
\end{eqnarray}
and the dispersion relation is in this case:
\begin{equation}
\left|\matrix{(1 - 4 \pi G x) &4 \pi G z\cr 4 \pi G z 
&(1 - 4 \pi G y)\cr}\right| = 0 .
\label{p11}
\end{equation} 
It can be shown that $\omega = 0$ is solution of that equation and correspond 
to the Goldstone boson corresponding to the breaking of the number of particle 
symmetry. 
Once we have obtained the energies $\omega_n$ of the different RPA roots, 
we can write the components of the RPA phonon in the form:
\begin{eqnarray}
 X_{n}(j,j) &=& \frac{4\pi G M(j,j)}{E_{j}+E_{j}-\omega_{n}}
\left(U_{j}^{2} + V_{j}^{2}\frac{4 \pi G z}{(1-4 \pi G y)}\right)
\Lambda_n \nonumber\\
 Y_{n}(j,j) &=& \frac{4\pi G M(j,j)}{E_{j}+E_{j}+\omega_{n}}
\left(U_{j}^{2}\frac{4 \pi G z}{(1-4 \pi G y)} + V_{j}^{2}\right)\Lambda_n ,
\nonumber \\
~
\label{p12}
\end{eqnarray}
where $\Lambda_n$ is determined by normalizing the phonon corresponding to the
$n-$th root of the RPA. The normalization condition reads
\begin{equation}
\sum_{j } [X_{n}^{2}(j) -Y_{n}^{2}(j)] = 1 .
\label{p13}
\end{equation}
Finally, we can obtain for each state $n$ the pairing strength parameter 
$\beta_{P}$ with the following formulae:
\begin{eqnarray}
\beta_{P}(2p) = \sum_{j} \sqrt{2j+1} \langle n|[a^{\dagger}_{j}
a^{\dagger}_{j}]_{00}|0\rangle = \nonumber \\
= \sum_{j} \sqrt{2j+1} [U^{2}_{j} X_{n}(j) - V^{2}_{j}Y_{n}(j)]\nonumber ,\\
\beta_{P}(2h) = \sum_{j} \sqrt{2j+1} \langle n|[a_{j}a_{j}]_{00}|0\rangle 
= & \nonumber \\ 
=\sum_{j} \sqrt{2j+1} [V^{2}_{j} X_{n}(j) - U^{2}_{j}Y_{n}(j)] .
\label{p14}
\end{eqnarray}
The predictions of the pairing strength distribution for the superfluid system 
$^{116}$Sn are shown in the two panels of Fig. 2.
For the calculation we have used the single-particle levels from Ref. \cite{US}. 
These last ones have been proved to give good results in BCS calculations using 
a pairing strength $G=g/A$, where $g\simeq 20 MeV$. We assume that the rest of 
the levels have occupation probability 1(0) if they are far below(above) the
Fermi surface. The change of the single particle energies around the Fermi 
surface has been done, in both cases, taking care of keeping the 
energy-centroids of the exchanged levels in the same position.  
The figure clearly shows the occurence of high-lying strength which can be 
associated with the Giant Pairing Vibration. Note that,with respect to the 
case of $^{208}$Pb, there is a minor fragmentation of the strength both in 
the low-lying and in the high-lying energy region.
\begin{figure}[!h]
\resizebox{1.0\hsize}{!}{\includegraphics*{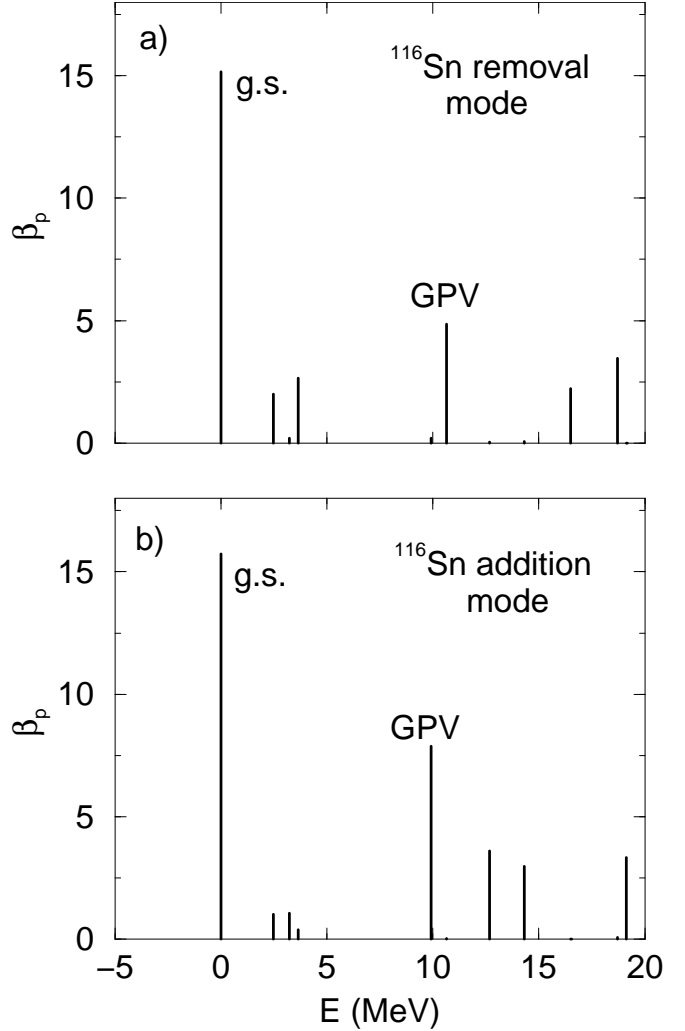}}
\caption{Pairing response for the removal (upper panel) and addition 
(lower panel) mode in $^{116}$Sn. The ground state transition and the 
candidate for GPV are evidenced.}
\end{figure}

\section{Macroscopic form factors for two-particle transfer reactions.}
The description of the reaction mechanism associated with the transfer 
of a pair of particles in heavy ion reactions has always been a rather 
complex issue.  In the limit in which the field responsible for the 
transfer process is the one-body field generated by one of the partners 
of the reactions, at least for simple configurations the leading order 
process is the successive transfer of single particles.  
In this framework the collective  
features induced by the pairing interaction arise from the coherence 
of different paths in the intermediate (A+1 , A--1) channel due to the 
correlation present in the final (A+2) and (A--2) states.  The actual 
implementation of such a scheme may turn out not to be a simple task, 
due to the large number of active intermediate states, and the use of 
a simpler approach may be desirable.  This is offered, for example, 
by the ``macroscopic model'' for two-particle transfer reactions, that 
parallels the formalism used to describe the inelastic excitation  
of collective surface modes.  In that case, as an alternative to 
the (more correct) 
microscopic description based on a superposition of particle-hole 
excitations, one has traditionally resorted to collective form factors  
of the form \cite{Sch} 
\begin{equation} 
F_{\lambda}(r)~=~\beta_{\lambda}R{dU \over dr} ,
\end{equation} 
in terms of the radial variation of 
the ion-ion optical potential U induced by the surface vibrations,  
with the strength parameter 
$\beta_{\lambda}$ obtained from the strength of the B(E$\lambda$)  
transition.  In the case of the pair transfer, the corresponding  
vibration is the fluctuation of the Fermi surface with respect to the 
change in the number A of particles, and the corresponding form 
factor $F_{P}$ is assumed to have the parallel form \cite{Dal}
\begin{equation} 
F_{P}(r)~=~\beta_{P}{dU \over dA} ,
\end{equation} 
in terms of the ``pairing deformation'' parameter $\beta_{P}$ 
associated with that particular transition, defined in the previous  
section. The assumption of simple scaling law between nuclear radius R and  
mass number A allows to rewrite the two-particle transfer form 
factor into an expression which is formally equivalent to the one  
for inelastic excitation, namely 
\begin{equation} 
F_{P}(r)~=~{\beta_{P} \over 3A}~R~{dU \over dr} .
\end{equation} 
This formalism has been successfully applied to quite a number of two-particle 
transfer reactions \cite{DP,DV}.  As in the case of inelastic excitations, macroscopic 
collective form-factors may in some cases only give a rough estimate 
to the data, requiring more elaborate microscopic descriptions.  Nonetheless, 
the use of simple macroscopic form factors is of unquestionable usefulness 
in making predictions, in particular in cases, as the one we are discussing,  
where experimental data are not yet available and estimates are needed in 
order to plan future experiments. 
 
\section{Applications: estimates of two-neutron transfer cross 
sections.} 
 
In order to evidence the possible role of unstable beams in the study 
of high-lying pairing states, we compare in this section two-particle transfer
reactions induced either by a traditionally available beam (e.g. the 
($^{14}$C, $^{12}$C)) or by a more exotic beam (e.g. the reaction 
($^{6}$He, $^{4}$He)).  As a target, we have considered the two cases of 
$^{208}$Pb and $^{116}$Sn, as representative cases of normal and superfluid  
systems in the pairing channels.  In both cases, we have considered the full 
pairing L=0 response, e.g. all transitions to 0$^+$ states in 
$^{210}$Pb and $^{118}$Sn, as described in Sect 2.
The Q-values corresponding to the transitions to the ground-states and to the 
GPV states are displayed in Table 1.
\begin{table}[!h]
\begin{center}
\begin{tabular}{c|c|c}
~~&~~$^{14}\mbox{C} \rightarrow ~^{12}\mbox{C}$~~&~~$^{6}\mbox{He}
 \rightarrow ^{4}\mbox{He}$\\ \hline
$^{116}\mbox{Sn} \rightarrow ~^{118}\mbox{Sn}_{gs}$~~& 3.15 MeV & 15.298 MeV\\
$^{208}\mbox{Pb} \rightarrow ~^{210}\mbox{Pb}_{gs}$~~& -4 MeV & 8.148 MeV\\
$^{116}\mbox{Sn}\rightarrow ~^{118}\mbox{Sn}_{GPV}$~~& -6.746 MeV & 5.402 MeV\\
$^{208}\mbox{Pb} \rightarrow ~^{210}\mbox{Pb}_{GPV}$~~& -15.81 MeV & -3.662 MeV
\end{tabular} 
\end{center}
\caption{Q-values for ground-state and GPV transitions. 
The target (column) and projectile (row) are specified.}
\end{table}  
For each considered state the two-particle transfer cross section has been 
calculated on the basis of the DWBA (using the code Ptolemy \cite{COM}) 
employing the macroscopic form factor described above, with a strength 
parameter as resulting from the RPA calculation. 
For the ion-ion optical potential, the standard parameterization 
of Akyuz-Winther \cite{Aky} has been used for the real part, with an imaginary part with 
the same geometry and half its strength.  
In all cases, the bombarding energy has been chosen in order to correspond, 
in the center of mass frame, to about 50\% over the Coulomb barrier. 
\begin{figure}[!t]
\resizebox{1.0\hsize}{!}{\includegraphics*{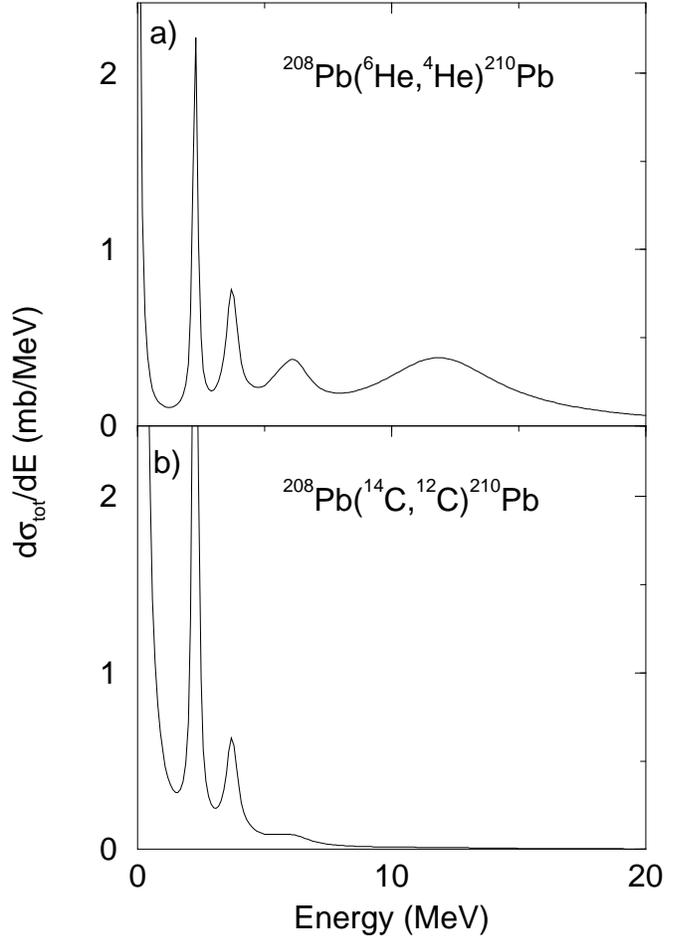}}
\caption{ Differential cross-sections as a function of the excitation energy
 for the two reactions : a) 
$^{208}$Pb($^{6}$He,$^{4}$He)$^{208}$Pb,
 and b) $^{208}$Pb($^{14}$C,$^{12}$C)$^{208}$Pb. See text for details.}
\end{figure}
The angle-integrated L=0 excitation function is shown in Fig. 3b as a 
function of the excitation energy E$_x$ for the  
$^{208}$Pb($^{14}$C,$^{12}$C)$^{208}$Pb reaction at E$_{cm}$=95 MeV.  For 
a more realistic display of the results, the contribution of each 
discrete RPA state is distributed over a lorentzian with $\Gamma$= k E$_x^2$,
with $k$ adjusted to yield a width of $4 MeV$ for the giant pairing vibration.
As the figure shows, the large (negative) Q-value associated with the region 
of the GPV (see Table 1) completely damps its contribution, and the excitation 
function is completely dominated by the transition to the ground state and the 
other low-lying states.  The situation is very different for the 
$^{208}$Pb($^{6}$He,$^{4}$He)$^{208}$Pb reaction at E$_{cm}$=41 MeV, 
whose excitation function is shown in Fig. 3a. In this case the weak binding 
nature of $^{6}$He projectile leads to a  
mismatched (positive) Q-value for the ground-state transition  
(Q$_{gs}$= 8.148 MeV), favouring the transfer process to the high-lying 
part of the pairing response.  In this case the figure shows that, 
in spite of a smaller pairing matrix element, the 
transition to the GPV is of the same order of magnitude of the 
ground-state transfer (1.8 mb for g.s. and 3.1 mb for the GPV). 
Note that a total cross section to the GPV region of 
the order of some millibarn should be accessible with the 
new large-scale particle-gamma detection systems.
\begin{figure}[!h]
\resizebox{1.0\hsize}{!}{\includegraphics*{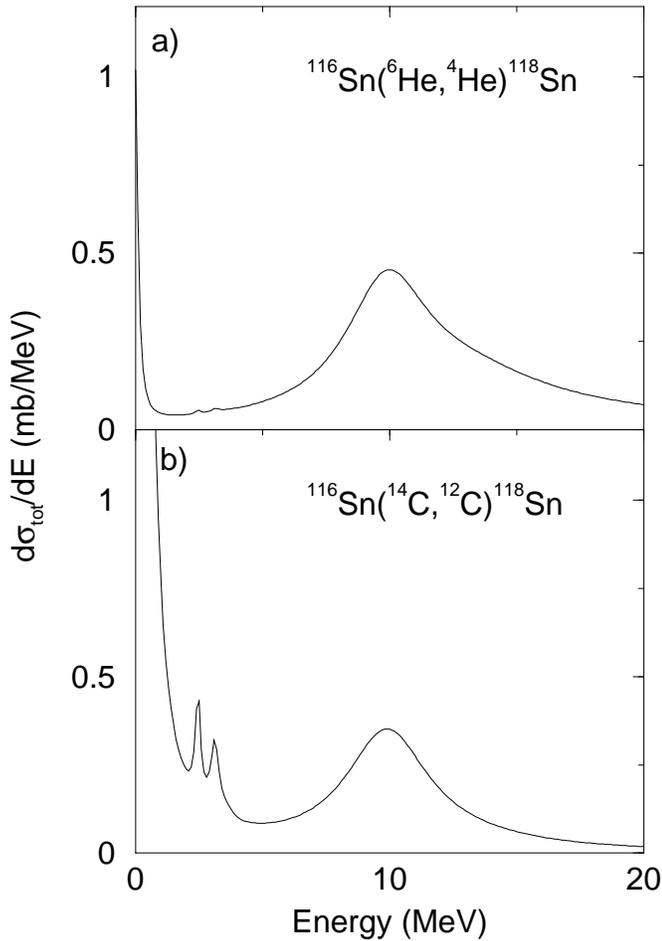}}
\caption{ Differential cross-sections as a function of the excitation energy
for the two reactions : a) 
$^{116}$Sn($^{6}$He,$^{4}$He)$^{118}$Sn,
and b) $^{116}$Sn($^{14}$C,$^{12}$C)$^{118}$Sn. The comparison between
the GPV and the ground-state  clearly shows the different strength.
 Notice the different vertical scale with respect to figure 3.  }
\end{figure}

A similar behaviour is obtained in the case of a tin target.
In Figs. 4a and 4b the corresponding excitation functions for the  
 $^{116}$Sn($^{14}$C,$^{12}$C)$^{118}$Sn reaction (at E$_{cm}$=69 MeV) 
and the $^{116}$Sn($^{6}$He,$^{4}$He)$^{118}$Sn reaction (at E$_{cm}$=40 MeV) 
are compared.Now the transition to the GPV dominates over the ground-state 
transition when using an He beam ( 0.4 mb for g.s. and 2.4 mb for the GPV). 
From a comparison with the RPA strength distributions of Fig. 1 and 2 one can 
see that the giant pairing vibrations is definitely favoured by the use of an 
$^{6}$He beam instead of the more conventional $^{14}$C one, because the 
transition to the ground-state is hindered, while the GPV is enhanced 
(or not changed), because of the effect of the Q-value.  

\section{Conclusions.}
The role of radioactive ion beams for studying different features of the pairing
degree of freedom via two-particle transfer reactions  is 
underlined. A $^6$He beam may allow an experimental study of high-lying 
collective pairing states, that have been theoretically predicted,
but never seen in measured spectra, because of previously unfavourable 
matching conditions.
The modification in the reaction Q-value, when passing from $^{14}$C to $^6$He, 
that is a direct consequence of the weak-binding nature of the latter 
neutron-rich nucleus, is the reason of the enhancement of the transition to the 
giant pairing vibration with respect to the ground-state.

\end{document}